\documentclass[fleqn, usenatbib, nofootnoteinbib]{mnras}
\usepackage[utf8]{inputenc}  
\usepackage[T1]{fontenc}
\usepackage{graphicx,psfrag}
\usepackage{mathrsfs}
\usepackage{amsmath,amsfonts,amssymb,pifont,gensymb}
\usepackage{multirow,enumerate}
\usepackage{comment,hyperref}
\usepackage{color}
\usepackage{acronym}
\usepackage{xspace}
\usepackage[normalem]{ulem}
\usepackage{mathtools}
\usepackage{makecell}
\usepackage{newtxtext,newtxmath}
\usepackage{changepage}
\usepackage{appendix}
\usepackage{lipsum}
\usepackage{tabularx}
\usepackage{subfig}
\usepackage{natbib}
\usepackage{float}
\usepackage{caption}

\usepackage{geometry}
\usepackage[dvipsnames]{xcolor}

 \newcommand{\lenshypo}{\mathcal{H}_\mathrm{L}}
 \newcommand{\unlenshypo}{\mathcal{H}_\mathrm{U}}
 \newcommand{\mods}{\mathcal{M}}
 \newcommand{\rgalsub}{\mathcal{R}_{\mathrm{gal}}^{\mathrm{sub}}}
 \newcommand{\blumc}{\mathcal{B}_{\mathcal{M}_c}}

\title[Uncovering faint lensed GWs with detected ones]{
Uncovering faint lensed gravitational-wave signals and reprioritizing their follow-up analysis using galaxy lensing forecasts with detected counterparts}
\author[L. Ng et al.]{
Leo C. Y. Ng$^{1}$\thanks{E-mail: leo.ng@ligo.org},
Justin Janquart$^{2, 3, 4, 5}$\thanks{E-mail: j.janquart@uu.nl},
Hemantakumar Phurailatpam$^{1}$,
\newauthor
Harsh Narola$^{2, 3}$,
Jason S. C. Poon$^{1}$,
Chris Van Den Broeck$^{2, 3}$,
Otto A. Hannuksela$^{1}$
\\
$^{1}$ Department of Physics, The Chinese University of Hong Kong, Shatin, N.T., Hong Kong \\
$^{2}$Institute for Gravitational and Subatomic Physics (GRASP), Department of Physics, Utrecht University, Princetonplein 1, 3584 CC Utrecht, The Netherlands \\
$^{3}$Nikhef – National Institute for Subatomic Physics, Science Park, 1098 XG Amsterdam, The Netherlands \\
$^{4}$ Center for Cosmology, Particle Physics and Phenomenology - CP3, Universit\'e Catholique de Louvain, Louvain-La-Neuve, B-1348, Belgium \\
$^{5}$ Royal Observatory of Belgium, Avenue Circulaire, 3, 1180 Uccle, Belgium
}

\date{\today}

\pubyear{2023}

\begin{document}

\maketitle

\begin{abstract}
    \noindent 
    Like light, gravitational waves can be gravitationally lensed by massive astrophysical objects. Strong gravitational lensing by galaxies and galaxy clusters is anticipated to become observable in the coming years. This phenomenon will manifest as multiple copies of the original wave, each exhibiting identical frequency evolution but distinct arrival times, amplitudes, and overall phases. Some of these images can be below the detection threshold and require targeted search methods, based on tailor-made template banks. These searches can be made more sensitive by using our knowledge of the typical distribution and morphology of lenses to predict the time delay, magnification, and image-type ordering of the lensed images. Here, we show that when a subset of the galaxy lensed images is super-threshold, they can be used to construct a more constrained prediction of the arrival time of the remaining signals, enhancing our ability to identify lensing candidate signals. Our suggested method effectively reduces the list of triggers requiring follow-up and generally re-ranks the genuine counterpart higher in the lensing candidate list. So, using information provided by the two or three super-threshold images, one can identify additional lensed images, also strengthening the evidence for the lensed signal hypothesis.
\end{abstract}


\section{Introduction}
Gravitational waves (GWs) were predicted by general relativity more than a hundred years ago~\citep{Einstein1916_GWs}. 
They are ripples in space-time propagating through the Universe and require cataclysmic phenomena to produce detectable signals. 
After its third observation run, the LIGO-Virgo-KAGRA (LVK) Scientific collaboration detected about a hundred mergers from compact binary coalescences, originating from binary black holes (BBHs), binary neutron stars (BNSs), and neutron star black hole (NSBH) binaries~\citep{LIGOScientific:2021djp}. 
These mergers were observed by the Advanced LIGO~\citep{TheLIGOScientific:2014jea} and Advanced Virgo~\citep{TheVirgo:2014hva} detectors. 
These observations have also enabled scientists to study black hole and neutron star populations~\citep{2021arXiv211103634T}, cosmological expansion~\citep{2021arXiv211103604T}, and general relativity in the strong field limit~\citep{2021arXiv211206861T}, among many other things. 
In addition, plans are ongoing to expand detection capabilities, with the upgrade of existing detectors and the development of new detectors, such as KAGRA~\citep{Somiya:2011np, Aso:2013eba, Akutsu:2018axf, Akutsu:2020his, Abbott:2020qfu} and LIGO-India~\citep{LigoIndia}. 
The increase afforded by more sensitive detectors will lead to an increased detection rate and has the potential to open the door to further scientific opportunities.
One such opportunity is the detection of lensed GW signals. 

Similarly to light, a massive object along a GW's travel path can lead to gravitational lensing~\citep{Ohanian1974, Degushi1986, Wang:1996as, Nakamura1998, Takahashi:2003ix}. 
Depending on the lens-source geometry and the mass of the lens, one can distinguish different types of lensing. 
For more massive lenses and good alignment, geometric optics applies, and the GW is split into multiple individually resolvable copies with the same frequency evolution, but magnified, time delayed, and potentially undergoing an overall phase shift~\citep{Dai:2017huk, Ezquiaga:2020gdt}. 
This is called strong lensing, typically produced by galaxy or galaxy cluster lenses. 
For galaxies, these signal copies arrive minutes to months apart~\citep{Dai:2017huk, Ng:2017yiu, Li:2018prc, Oguri:2018muv, Wierda:2021upe}; for galaxy clusters, the copies might arrive up to years or more apart~\citep{Smith:2017mqu, Smith:2018gle, Smith:2019dis, Robertson:2020mfh, Ryczanowski:2020mlt}. 
When the lens is less massive, one might observe so-called millilensing~\citep{Liu:2023ikc}. 
This is still the geometric optics regime but the time delay is much shorter and the various images overlap in the detector band, leading to a single non-trivial GW. 
For even lighter lenses, the GWs enter the wave optics regime, leading to frequency-dependent beating patterns~\citep{Takahashi:2003ix, Cao:2014oaa, Lai:2018rto, Christian:2018vsi, Singh:2018csp, Meena:2019ate, Pagano:2020rwj, Cheung:2020okf, Kim:2020xkm, Wright:2021cbn}, sometimes referred to as microlensing. 
Over the last years, searches for lensing have been carried out, but no significant evidence has been found so far~\citep{Hannuksela:2019kle, McIsaac:2019use, Li:2019osa, Dai:2020tpj, LIGOScientific:2021izm, LIGOScientific:2023bwz, Janquart:2023mvf}. 

In this work, we focus on GWs stronlgy lensed by galaxies. 
Several independent works have predicted such events to become detectable in the coming years~\citep{Ng:2017yiu, Li:2018prc, Oguri:2018muv, Wierda:2021upe, Xu2021PleasePopulations}. 
Strong lensing by galaxies would lead to two to four potentially detectable images~\citep{Dai:2017huk, Ezquiaga:2020gdt, Liu:2020par, Lo:2021nae, Janquart:2021qov}. 
To search for such events, one attempts to identify GWs with a similar frequency evolution~\citep{Haris:2018vmn, Liu:2020par, Lo:2021nae, Janquart:2021qov, Janquart:2023osz, Goyal:2020bkm}. One method of searching for lensed events is by considering GW events already detected using already identified events (super-threshold events)~\citep{Hannuksela:2019kle, LIGOScientific:2021izm, LIGOScientific:2023bwz, Janquart:2023mvf}. 
However, such a search can potentially miss interesting candidates. 
In particular, the (de)magnification of images and the change in observation conditions due to the time delay and Earth's rotation means that some images can exist below the noise floor as sub-threshold candidates~\citep{Li:2019osa, McIsaac:2019use, Dai:2020tpj, Li:2023tex}. 
If such sub-threshold candidates are identified, it could lead to an increased confidence in detections and lead to an improved strong lensing science case. To target such events, the seminal approach is to make use of the super-threshold events to perform an additional search using template banks (see~\citet{Messick_2017, Sachdev:2019vvd, Hanna_2020, cannon2020gstlal, Adams_2016, Aubin_2021, Allen_2012, Allen_2005, Dal_Canton_2014, Usman:2015kfa, Nitz_2017, Davies_2020} for details on template bank searches) that target only event triggers with waveforms similar to the super-threshold waveforms~\citep{Li:2019osa, McIsaac:2019use, Dai:2020tpj}. 
Such sub-threshold lensing searches have already been used to search for lensed counterparts of LVK signals~\citep{McIsaac:2019use, LIGOScientific:2021izm, LIGOScientific:2023bwz, Janquart:2023mvf}.
The output of the search is a list of triggers to be followed up using usual analysis methods (see~\citet{LIGOScientific:2023bwz} for an example). 

However, the number of potential candidate triggers scales quadratically with time and can thus be computationally expensive. 
Furthermore, the list obtained is often composed of many candidates also coming from noise artifacts or unrelated signals. 
Therefore, it is important to identify methods that reduce false noise triggers and computational costs. 

In~\citet{Goyal:2023lqf}, the authors utilised strong galaxy lensing forecasts to re-rank candidate triggers for follow-up analysis; an advantage offered by galaxy strong lensing is that we have good models for this situation and we can compute the expected distributions for the time delay and relative magnification of the events~\citep{Haris:2018vmn, Wierda:2021upe, More:2021kpb}. 
The idea is then that those candidates with time delays consistent with strong lensing predictions should be ranked higher than those that do not. 

In this work, we consider a scenario where we have observed two or three super-threshold images. 
In particular, if a subset (for example 3 out of 4 images) of the GW images have already been identified using super-threshold searches, our predictive power for the time delays increases. 
Thus, we simulate several different scenarios in which a subset of the images have already been detected and explore how predicting the time delay for the corresponding sub-threshold images can improve the ability to identify additional lensing candidates. 

This work is structured as follows. 
In Sec.~\ref{sec:strong_lensing}, we describe strong lensing of GWs, focusing on galaxy lensing. 
Then, we give an overview of the current lensing searches for super- and sub-threshold events in Sec.~\ref{sec:overview_lensing_searches}. 
Next, we describe how one can construct an adapted probability density for the time delay of the sub-threshold image based on the observed super-threshold ones in Sec.~\ref{sec:constructing_kde}. 
We illustrate the method in Sec.~\ref{sec:results}. 
Finally, we give our conclusions and future perspectives in Sec.~\ref{sec:conclusions}.

\section{Gravitational Wave Strong Lensing by Galaxies}
\label{sec:strong_lensing}

\begin{figure}
    \centering
    \includegraphics[width=\linewidth]{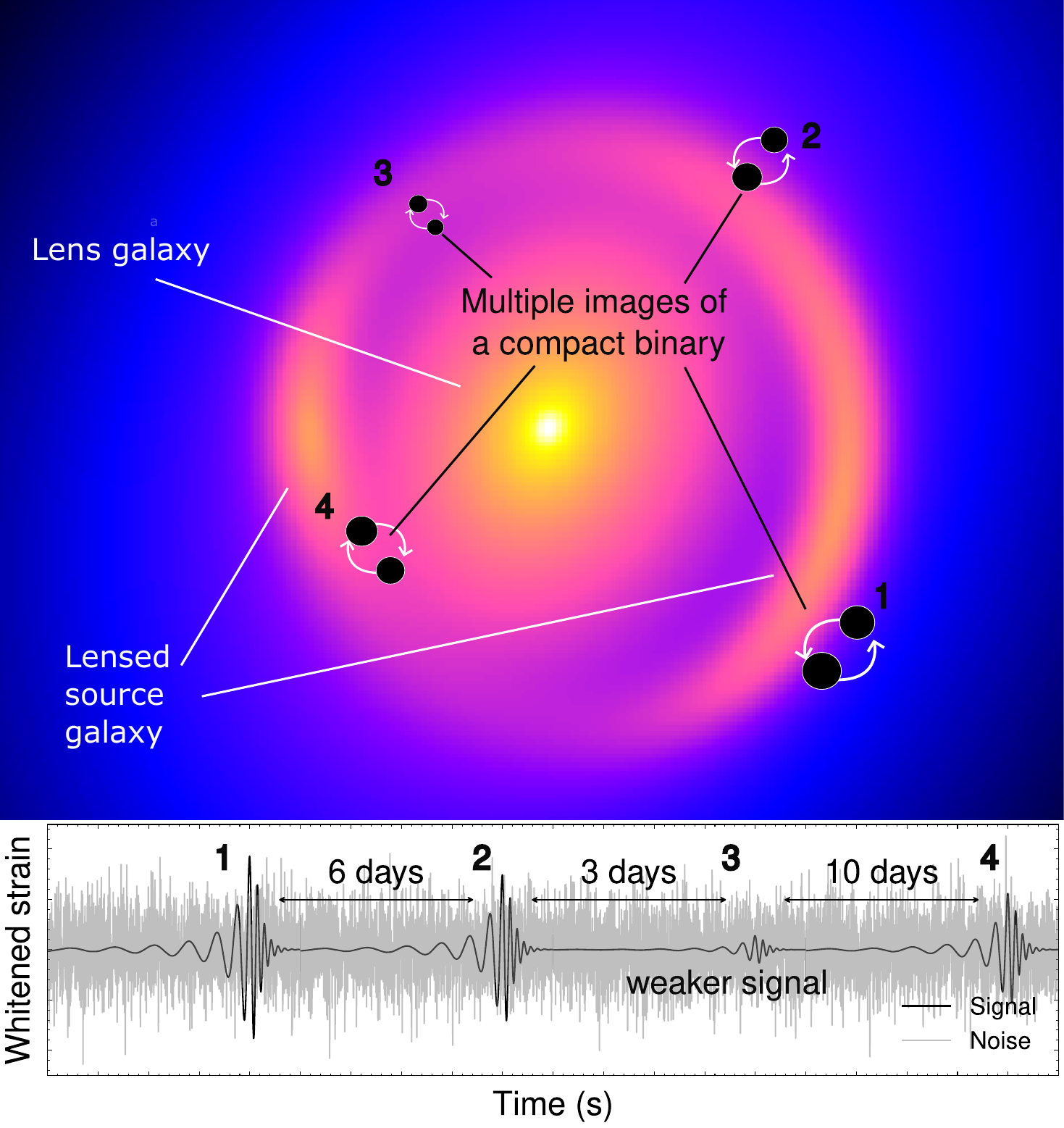}
    \caption{Illustration of an example of a strong lensing scenario. A galaxy lens strongly lenses a compact binary object and produces four potentially observable images that appear in the GW detectors as multiple copies of the same signal. 
	These copies appear otherwise identical but differ in their arrival times, amplitudes, and can potetially undergo an overall phase shift, due to lensing time delay, magnification, and possible overall discrete phase shifts, respectively. 
	A typical galaxy lens produces either three or five images of the binary compact object, but one of the central images is typically heavily demagnified such that it is neglected and only two or four images are observed, with the two type-I images arriving first and the two type-II images arriving later. 
	Since the amplitude and arrival time of the signal copies are different, some of the signals may reside below the noise threshold. 
	Significant effort has been taken to dig such signals out of the noise background using searches with template banks tailored to be more sensitive to lensing. 
	One method to improve the sensitivity of these searches is by incorporating knowledge about the distribution and morphology of galaxy lenses to predict where the signals should appear in the data.
    }
    \label{fig:illustration_lensing}
\end{figure}

When strong lensing occurs, GWs are split into multiple distinct, potentially detectable, copies with the same frequency evolution\footnote{When higher-order modes are present and lensing leads to a specific extra phase shift, the frequency evolution can slightly be affected.}---referred to as \emph{images}~\citep{Takahashi:2003ix}. 
Compared to unlensed signals, the lensed images undergo an overall (de)magnification, a time delay, and an overall phase shift. 
The waveform for the $j^{th}$ lensed image ($h^{j}_{\mathrm{L}}$) is related to the unlensed waveform ($h_{\mathrm{U}}$) as~\citep{Dai:2017huk, Ezquiaga:2020gdt}
\begin{equation}\label{eq:lensed_unlensed_gw}
\tilde{h}_L^j (f; \vec{\theta}, \vec{\Lambda}_j) = \sqrt{\mu_j}\,e^{(2\pi i f t_j - i \pi n_j \text{sign}(f))}\, \tilde{h}_U (f; \vec{\theta})\,,
\end{equation}
where the tilde denotes the frequency domain, $\vec{\theta}$ represents the usual BBH source parameters, and $\vec{\Lambda}_j$ represents the lensing parameters ($\vec{\Lambda}_j = \{\mu_j, t_j, n_j\}$, with $\mu_j$ the magnification changing the amplitude of the wave, $t_j$ the extra time delay due to the change in geometrical path and the Shapiro delay~\citep{Shapiro:1964gre}, and $n_j$ the Morse factor leading to an overall phase shift). 
Fig.~\ref{fig:illustration_lensing} shows a representation of a lensed quadruplet.

Eq.~\eqref{eq:lensed_unlensed_gw} is obtained by considering the stationary points of the time delay function like Fermat's principle~\citep{Takahashi:2003ix}. 
Therefore, only particular points are a solution, and each of these points gives a specific value for the Morse factor. 
A lensed image can be classified into different image types according to its Morse factor. Type I images have $n_j = 0$, corresponding to a minimum of the Fermat potential. 
This is equivalent to having no phase shift at all. 
When $n_j = 0.5$, the image is type II and comes from a saddle point of the potential. 
Finally, when the image corresponds to a maximum of the potential, we have a type III image and $n_j = 1$. 
Type I and III images are indistinguishable as one can be transformed into the other by applying a $\pi/2$ change in the GW polarization, which is undetectable~\citep{Ezquiaga:2020gdt}. 
In contrast, for type II images, the lensing signature is detectable provided the higher-mode content is sufficiently large~\citep{Ezquiaga:2020gdt, Janquart:2021nus, Wang:2021kzt, 2022arXiv220206334V}. 

When a GW is lensed by a galaxy, we expect to see two to four images with a specific image-type ordering, with image types ordered as I--I--II--II ---where the last or last two type II images may not form for some lens-source configurations, and separated by days to months~\citep{Dai:2017huk, Ng:2017yiu, Li:2018prc, Oguri:2018muv, Ezquiaga:2020gdt, Wierda:2021upe, More:2021kpb}. 
In some rare cases with a very shallow profile, we can also have a faint but observable central type III image~\citep{2013ApJ...773..146D, Collett:2017ksf}. 
This ordering is valid only for galaxy lenses. 
Events strongly lensed by a galaxy cluster can have more complicated lensing morpologies~\citep{Smith:2017mqu, Smith:2018gle, Smith:2019dis, Robertson:2020mfh, Ryczanowski:2020mlt}. 
We do not consider the latter in this work. 
Thus, when a galaxy-lensed GW event is seen, we expect the type-I images to arrive first, followed by type-II signals. 

However, because of the change in observation condition from one image to the other---possible offline detectors from one image to the other or different sensitivity to that region of the sky because of Earth rotation---and possible demagnification relative to other images, all images are not always observable. 
In such scenarios, the observed lensed images can be used to predict when the other counterparts should be detected. 
Indeed, if we assume a super-threshold image is lensed and no super-threshold counterpart is found, then the lensing hypothesis predicts that there should exist other images that could potentially be found (see the ``hidden'' third gravitational wave in the illustrative Fig.~\ref{fig:illustration_lensing}). 

Observing multiple images of a lensed event can make the process of predicting the arrival time of the next image more effective. 
For instance, if three images are observed, rudimentary lensing theory necessitates that a fourth image must exist, even if it was not identified. 
Furthermore, the arrival time of the fourth image can be predicted using the observed arrival times of the other images quite accurately. 
As an illustrative example, we show the time-delay prediction using the software \texttt{LeR}~\citep{phurailatpam_hemantakumar_2023_8087461}. The latter produces expected lensing time-delay distributions of binary black holes sampled from the latest \texttt{PowerLaw+Peak} population model based on O3 data together with galaxy lenses sampled from the LSST galaxy catalogue (see Sec.~\ref{sec:constructing_kde} for details).
Similarly, for two images, the arrival time of the remaining images can be predicted, albeit with a larger uncertainty.
In this work, we investigate the use of such arrival time predictions to re-rank triggers coming from sub-threshold searches when at least two super-threshold events are observed.

Unfortunately, it is generally not possible to tell precisely which lensed image in the series was missed. 
That is, for example, in a four-image scenario with the third image having been missed (Fig.~\ref{fig:illustration_lensing}), we cannot tell if the missed image was the third or the fourth image. 
Instead, we can infer only a subset of information using the image types of the detected images. 
Since the image types are ordered (I--I--II--II), we can infer that the missed image is of type II, but we cannot tell if it was the third or the fourth such image. 

Furthermore, only the Morse factor difference of the missed image can be inferred from the observed images because it is often degenerate with the coalescence phase of the waveform~\citep{Ezquiaga:2020gdt}.\footnote{An exception is that the Morse factor can be measurable for type II images with a large higher-order mode content~\citep{Janquart:2021nus, Wang:2021kzt, 2022arXiv220206334V}. } 
Similarly, the magnification and time delay are fully degenerate with the luminosity distance and the coalescence time of the wave, respectively~\citep{Dai:2017huk, Ezquiaga:2020gdt}. 
Therefore, it is common to work with the relative lensing parameters (relative Morse phases, magnifications, and arrival times) between two images instead of the absolute parameters (absolute Morse phases, magnifications, and arrival times) (e.g.~\citet{Lo:2021nae, Janquart:2021qov}). 

The relative lensing parameters between an image $j$ and another image $l$ are denoted $\vec{\Phi}_{lj} = \{\mu_{lj}, t_{lj}, n_{lj}\}$ with the relative magnification $\mu_{lj} = \mu_{l}/\mu_j$, the time delay $t_{lj} = t_l - t_j$, and the Morse factor difference $n_{lj} = n_l - n_j$. 
Naturally, since the individual Morse factors are discrete, their differences also are. 
In the end, the Morse factor difference---which is generally recovered~\citep{Janquart:2021qov, Janquart:2023osz}---does not give direct access to the individual Morse factors. However, the GW observation gives access to the Morse factor \emph{difference} between signals, which is sufficient to improve the arrival time prediction for the next image.
The key assumption is that one can predict the image arrival time ordering (type I--I--II--II). 
For example, a Morse factor difference of zero for two detected images implies that the images are of the same type and are either the first two or the last two signals.
A Morse factor difference $n_{lj} = 0.5$ between two signals implies that one is type I and the other type II and they are any combination of the first two and last two signals \footnote{Strictly speaking, we could also have one is type II and the other type III but type III images are very rare and therefore we neglect them in this work.}. 
Table~\ref{tab:MorseFactorDiffs} gives a full list of Morse factor differences and their implications. 
In this work, we investigate the use of these measured relative lensing parameters to determine what characteristics are expected for any missed sub-threshold images. 
We detail our method in Sec.~\ref{sec:constructing_kde} after a brief description of current lensing searches in the next section.

\begin{table*}%[hb]
\begin{minipage}{\textwidth}
  \centering
  \begin{tabular}{|c|l|l|c|}
    \hline
    $n_{j1}$\footnote{
        Note that the number 1 in $n_{j1}$ and the number 1 in the central two columns may not refer to the same image if the first image of the lensed event is a sub-threshold event.
      } & Detected image(s) & Undetected image(s) & Total no. of sub-cases \\
    \hline
    / 
      & $\{1\}$ & \{2, 3, 4\} & 12 \\
      & $\{2\}$ & \{1, 3, 4\} & \\
      & $\{3\}$ & \{1, 2, 4\} & \\
      & $\{4\}$ & \{1, 2, 3\} & \\
    \hline
    0 
      & $\{1, 2\}$ & \{3, 4\} & 4 \\
      & $\{3, 4\}$ & \{1, 2\} & \\
    \hline
    $\frac{1}{2}$ 
      & $\{1, 3\}$ & \{2, 4\} & 8 \\
      & $\{1, 4\}$ & \{2, 3\} & \\
      & $\{2, 3\}$ & \{1, 4\} & \\
      & $\{2, 4\}$ & \{1, 3\} & \\
    \hline
    $\{0, \frac{1}{2}\}$
      & $\{1, 2, 3\}$ & \{4\} & 2 \\
      & $\{1, 2, 4\}$ & \{3\} & \\
    \hline
    $\{\frac{1}{2}, \frac{1}{2}\}$ 
      & $\{1, 3, 4\}$ & \{2\} & 2 \\
      & $\{2, 3, 4\}$ & \{1\} & \\
    \hline
  \end{tabular}
  \caption{Various sub-cases possible for quadruply lensed events when one, two, or three super-threshold images are observed. $n_{j1}$ is the relative Morse factor of the $j^{th}$ super-threshold image compared to the first super-threshold image. The first column gives the measured Morse factor difference for the detected images. The second and third columns list the different images that could have been detected and the corresponding possible sub-threshold ones, respectively. The last column gives the total number of sub-cases to consider for a given measured Morse factor difference.}
  \label{tab:MorseFactorDiffs}
\end{minipage}
\end{table*}

\section{Searches for Super- and Sub-Threshold Lensed Images}
\label{sec:overview_lensing_searches}

\subsection{Overview of Lensing Searches}

Typically, strong lensing searches for multiple images are done in two ways. 
The first focuses on the GW events detected with the unlensed searches. 
The idea is then to see if the pool of candidates contains event multiplets with similar frequency evolution, as expected from strong lensing. 
The identification is typically done in multiple steps, starting from the least expensive and less precise methods to the most precise and computationally heavier ones~\citep{Hannuksela:2019kle, LIGOScientific:2021izm, LIGOScientific:2023bwz, Janquart:2023mvf}. 
For the first filtering, two methods are used: posterior overlap~\citep{Haris:2018vmn}, where one tests for the compatibility between the posteriors of parameters unaffected by lensing, and \textsc{LensID}~\citep{Goyal:2020bkm}, a machine learning pipeline comparing the sky and time-frequency maps of the events forming the pair of interest. 
The GW pairs not discarded by these methods are then passed to the next step, done by \textsc{GOLUM}~\citep{Janquart:2021qov, Janquart:2023osz}, which uses the posteriors of the first image to analyze the strain of the second one. 
The remaining pairs are passed to a full joint Bayesian analysis framework also including selection and population effects to obtain a Bayes factor~\citep{Lo:2021nae}\footnote{Other codes implementing joint parameter estimation exist but they generally do not include the full population and selection effect procedure~\citep{Liu:2020par, Janquart:2023osz}. 
In this work, when doing joint parameter estimation, we use the \textsc{GOLUM} package~\citep{Golum_git}.}. 
However, if some of the images are not detected by the traditional searches---typically because they are too faint, these events will be missed.
Sub-threshold searches exist to remedy this problem. These target GW images that are counterparts of the detected GW events by constructing a reduced template bank~\citep{Li:2019osa, McIsaac:2019use, Dai:2020tpj, Li:2023tex}. 
%For example, searching for a sub-threshold candidate for a given high-mass binary black hole could be performed by searching the GW data with only waveforms with identical frequency evolution, instead of the entire bank of all possible waveforms. 
This reduces the bank's trial factor and enables one to recover fainter triggers compared to the full bank. 
However, such searches can lead to a relatively long list of triggers as they also pick up unrelated fainter signals or noise artefacts. 

\subsection{Improving the detection confidence of candidates using lensing}
\label{ssec:overview_lensing_searches}

It is possible, in principle, to improve one's ability to dig out such faint lensed signals from noise by predicting the arrival time of the signal. 
This is one of the key components of, for example, the so-called \texttt{PyGRB} search, which uses trigger times of gamma-ray bursts to associate signals within a given short time window~\citep{harry2011targeted, Williamson:2014wma}. 
In such a case, the arrival time is predicted to be around the GRB event.
In lensing, the arrival time is predicted using the relative lensing parameters of the detected images. 
Indeed, if the arrival time is known to have a higher precision, it will be less likely that a noise trigger will exist within that short period than a timespan of the entire dataset spanning a year or more. 

In more detail, from a Bayesian perspective, one can quantify the probability ratio of a set of data $\vec{d}$ containing a signal,
\begin{equation}
\begin{split}
  \frac{p(\mathcal{H}^\text{signal}|\vec{d})}{p(\mathcal{H}^\text{noise}|\vec{d})} 
  &= \frac{p(\vec{d}|\mathcal{H}^\text{signal})}{p(\vec{d}|\mathcal{H}^\text{noise})} \frac{p(\mathcal{H}^\text{signal})}{p(\mathcal{H}^\text{noise})} \\
  &= \mathcal{B}^\text{signal}_\text{noise} \frac{p(\mathcal{H}^\text{signal})}{p(\mathcal{H}^\text{noise})} \,,
\end{split}
\end{equation}
where $\mathcal{H}^\text{signal}$ and $\mathcal{H}^\text{noise}$ are the hypotheses that the data contain a signal or not, respectively. 
The first term on the right-hand side is the likelihood ratio, also called the Bayes factor ($\mathcal{B}^\text{signal}_\text{noise}$), and the second term is the prior odds.

Analogously, the probability that a set of data $\vec{d}$ contains a lensed signal is given by 
\begin{equation}
\begin{split}
  \frac{p(\mathcal{H}_L^\text{signal}|\vec{d})}{p(\mathcal{H}^\text{noise}|\vec{d})} 
  &= \frac{p(\vec{d}|\mathcal{H}_L^\text{signal})}{p(\vec{d}|\mathcal{H}^\text{noise})} \frac{p(\mathcal{H}_L^\text{signal})}{p(\mathcal{H}^\text{noise})} \,,
\end{split}
\end{equation}
where $\mathcal{H}_L^\text{signal}$ is the hypothesis that the data contains a lensed signal, $\vec{d}$ is the data, and $p(\vec{d}|\mathcal{H}_L^\text{signal})$ is the likelihood of the data given the lensed signal hypothesis. 
The prior odds (ratio of prior probabilities) are documented in (Hannuksela et al., in prep.).
The Bayes factor (ratio of evidences) can be related to the traditional lensing Bayes factor that \textsc{GOLUM} and other joint parameter estimation (JPE) pipelines compute as
\begin{equation}
\begin{split}
  \mathcal{B}^{\text{signal}, L}_\text{noise} 
  &= \frac{p(\vec{d}|\mathcal{H}_L^\text{signal})}{p(\vec{d}|\mathcal{H}_U^\text{signal})}
  \frac{p(\vec{d}|\mathcal{H}_U^\text{signal})}{p(\vec{d}|\mathcal{H}^\text{noise})} \\
  &= \mathcal{B}^L_U \times \mathcal{B}^\text{signal}_\text{noise} \,,
\end{split}
\end{equation}
where $\mathcal{B}^L_U$ is the Bayes factor between the lensed and unlensed signal hypotheses, and $\mathcal{B}^\text{signal}_\text{noise}$ is the Bayes factor between the unlensed signal and noise hypotheses. 
The Bayes factor contains both information regarding the waveform model and the arrival time of the signal. 
In particular~\citep{Haris:2018vmn},
\begin{equation}
  \mathcal{B}^L_U \propto \frac{p(\vec{t}|\mathcal{H}_L)}{p(\vec{t}|\mathcal{H}_U)} \,,
\end{equation}
where $\vec{t}$ is the arrival time of the signal. 
Therefore, a more informative arrival time prediction will allow one to be more confident that a signal candidate is a real signal and not noise, \emph{i.e.} $\mathcal{B}^{\text{signal}, L}_\text{noise}$ increases.

It is worth noting that care would need to be taken in implementing such arrival time predictions and additional lensing information in the sub-threshold searches in practice, which is beyond the scope of this work. 
The Bayes factor that is often cited in the context of lensing joint parameter estimation is derived after applying extensive Bayesian analyses methods. For searches, the information is found by looking through a template bank. Instead, one uses different statistics to describe the likelihood of the trigger being a GW signal against it being noise. This is based on a likelihood ratio for the GstLAL pipeline~\citep{Li:2019osa, Li:2023tex}, or other ranking statistics for the PyCBC ones~\citep{McIsaac:2019use, Nitz_2020, Davies_2020}. This statistic is the same between dedicated lensing searches and usual unlensed ones. We also note the possibility of having a difference in data quality between the data used for searches and parameter estimation analyses since more in-depth data cleaning is applied to the latter if needed. Adding lensing-based information will enhance the confidence in the lensing status of detected triggers.
For these reasons, this work investigates how much the arrival time will aid in identifying and re-ranking triggers to understand importance of an implementation of accurate arrival time predictions.  
The realistic implementation of this method in the searches is beyond the scope of this work. 
%The next section explains how we can build such a distribution and how it can be used to re-rank the sub-threshold triggers.

\section{Predicting Time Delays for Sub-Threshold Lensed Counterpart}
\label{sec:constructing_kde}

\subsection{The Lensed Time Delay Distribution}

Suppose one detects two or three super-threshold lensed images, with one or more sub-threshold images that have been missed by standard searches, to predict the arrival time of the sub-threshold images. 
In that case, one can analyze the super-threshold events using JPE~\citep{Liu:2020par, Lo:2021nae, Janquart:2021qov, Janquart:2023osz}. 
The procedure gives one access to 
(i) the Morse factor difference, generally seen as a discrete phase difference between two images, 
(ii) the difference in arrival times, obtained with millisecond precision, and
(iii) the relative magnifications, less precisely determined as their error are correlated with the luminosity distance uncertainty. 
Since the relative magnifications are subject to higher uncertainty, they tend to be less relevant in predicting the arrival times of the sub-threshold events.
Nevertheless, for completeness, we use all three relative lensing parameters to predict the time delay distributions. 

Using the Morse factor, arrival time, and relative magnification differences, the goal is to find the arrival time of any sub-threshold images in the data. 
That is, to forecast the arrival time of the sub-threshold event using the time-delay prior:
\begin{equation}\label{eq:dt_lenshypo_full}
  p(t_{s1} | \vec{d}_\text{super}) = \int p(t_{s1} | \vec{\Phi}_{j1}, \lenshypo, \mods) p(\vec{\Phi}_{j1}| \vec{d}_\text{super}, \lenshypo, \mods) \text{d} \vec{\Phi}_{j1} \,,
\end{equation}
where $\vec{d}_\text{super}$ are all the gravitational-wave data from the super-threshold events, $t_{s1}$ is the time delay between the sub-threshold trigger and the super-threshold image arriving first, $\vec{\Phi}_{j1}$\footnote{From here on, the sub-scripts correspond to the ordering of the \emph{detected} super-threshold images and not the position of the images in the lensed multiplet. So, for example, $\vec{\Phi}_{21}$ corresponds to the relative parameters between the first and second detected super-threshold images.} 
are the relative lensing parameters between the first and all other (second and/or third) super-threshold images---depending on the lens-source system---and $\mods$ represent the assumed BBH and lens models, which can be chosen when doing the analysis.
The posterior distribution $p(\vec{\Phi}_{j1}| \vec{d}_\text{super}, \lenshypo, \mods)$ are thus the measured lensing parameters from the super-threshold GW data. 

\begin{figure*}
    \centering
    \includegraphics[width=\textwidth]{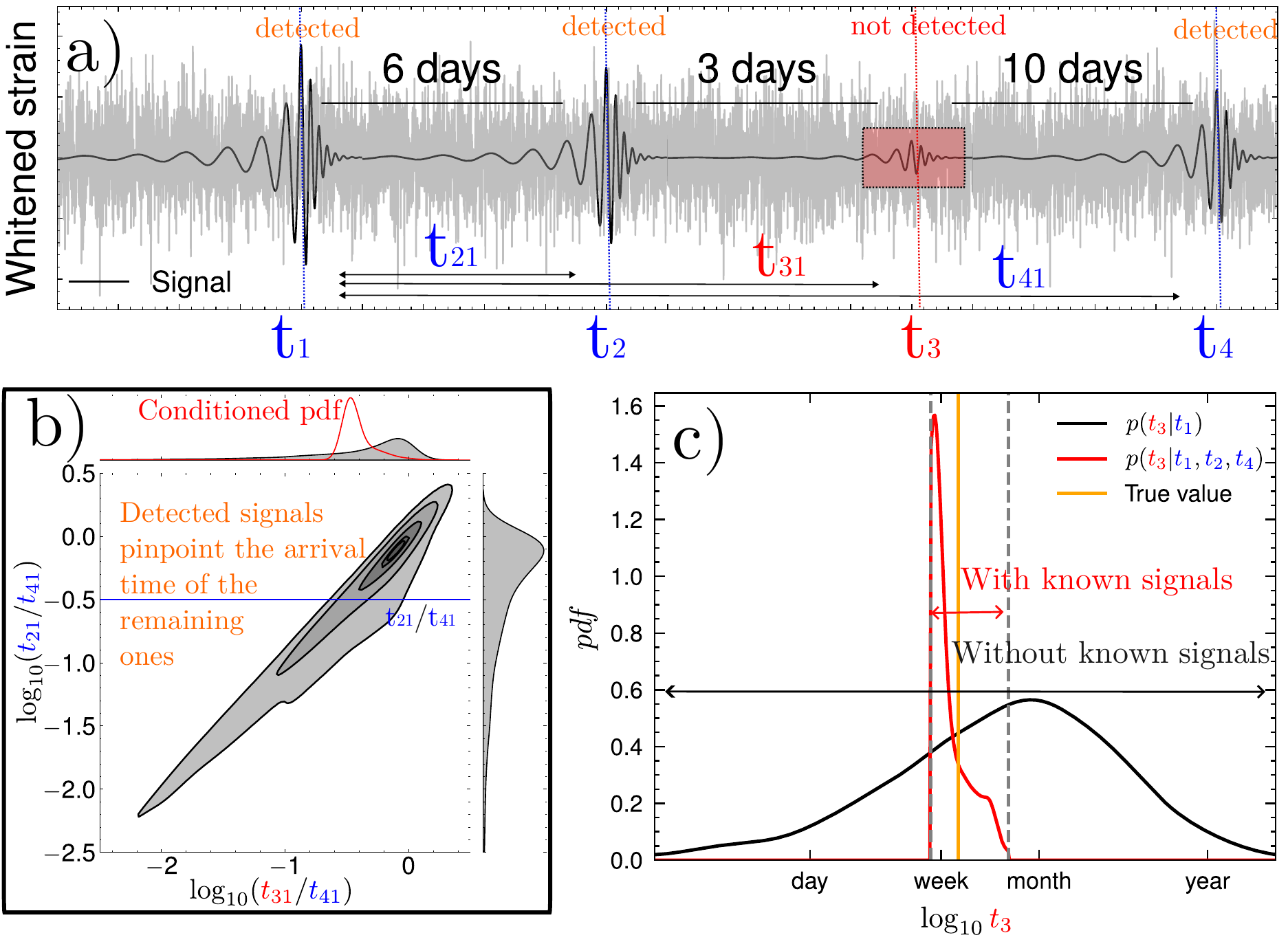}
    \caption{An example illustration of predicting the arrival time of a sub-threshold signal using three detected super-threshold signals. 
    The four images are separated by 6, 3, and 10 days, respectively, with the first two and the last image detected (panel a). 
    By using the knowledge of the arrival times of the detected signals, one can predict the arrival time of the remaining third signal (panel b). 
    The knowledge of the arrival improves one's ability to pinpoint the arrival time of the signal from a time window spanning over a year to a small time window of a week (panel c). 
    Here we assume that it is known that the third signal is missing. 
    However, in general, one would not be able to tell if the fourth detected image is the third or the fourth signal, which broadens the error on the time delay prediction (see Table~\ref{tab:MorseFactorDiffs}).
    Here the first signal arrives at $t_1=0$.}
    \label{fig:time_delay_illustration}
\end{figure*}

Since the GW arrival times are resolved at very high accuracy ($\sim$millisecond) compared to the lensing time delays ($\sim$hours to months), we take the arrival times of the super-threshold events to be a delta function. 
%The approximation is widely used in joint parameter estimation~\cite{Liu:2020par, Lo:2021nae, Janquart:2021qov, Janquart:2023osz} and is justified by the fact that the arrival time of the super-threshold events is well determined by the gravitational-wave data.
Furthermore, since the relative magnification plays a less important role in the time delay prediction, we take the relative magnifications to be likewise a delta function at a single value (the median). 
In compact notation, the time-delay prediction for the sub-threshold event becomes
\begin{equation}\label{eq:dt_lenshypo}
  p(t_{s1} | \vec{d}_\text{super}) = p(t_{s1} | \vec{\Phi}_{j1}, \lenshypo, \mods) \,,
\end{equation}
where now $\vec{\Phi}_{j1}$ are understood to be the median. 
From this point onwards, we will make this approximation and drop the explicit mention of the GW data and posterior, though the method can be applied to all points in the posteriors to obtained a final distribution as output. 
If we select a given trigger and evaluate Eq.~\eqref{eq:dt_lenshypo}, we obtain the probability to observe a trigger with a specific time delay given the characteristics of the super-threshold events and the assumed lens and source models. The process is illustrated in Fig.~\ref{fig:time_delay_illustration}.

To construct the time delay prediction (the probability density function described in Eq.~\eqref{eq:dt_lenshypo}), we rely on a Gaussian kernel density estimator (KDE) constructed based on a large catalogue of lensed events. 
The latter is generated using the \texttt{LeR} package~\citep{phurailatpam_hemantakumar_2023_8087461}, an efficient code to generate lensing statistics. 
Principally, the code generates the catalogue by sampling the characteristics of the sources and lenses from chosen source and lens populations, then solving the lens equation and retaining only the setups leading to multiple images. 
Here, we use the \texttt{PowerLaw+Peak} BBH mass model~\citep{2021arXiv211103634T}, the redshift distribution from~\citet{Oguri:2018muv}, and an elliptical power-law with shear lens model~\citep{Tessore_2015}. 
For all the models, we use the same parameters as the one specified in~\citet{Wierda:2021upe}. The parameters are also given in Appendix~\ref{appenB}.
Here, and throughout the work, we use the detector network made of the two LIGO and the Virgo detectors and assume they are at their expected O4 sensitivity~\citep{Abbott:2020qfu}. 
When selecting the samples on which we apply the KDE, we consider as those with a network signal-to-noise-ratio (SNR) higher than 8 super-threshold events and those with a network SNR between 6 and 8 as sub-threshold. 
The value of 8 is chosen because, in Gaussian stationary noise, it is a good proxy for detectability~\citep{Essick:2023toz}. 
For the sub-threshold case, the SNR values corresponding to identifiable signals are more uncertain, but here, we take an SNR cut at 6 to remove the faintest signals, which would be very hard to detect even with adapted search methods. 
Using \texttt{LeR}, we generate a catalogue containing $2\times10^6$ lensed systems. 
This number is chosen to ensure having enough samples representing the possible lensing scenarios while keeping a reasonable generation time for the catalogue.

% Original Table 1

Once we have a catalogue at our disposal, we can construct the KDE for a given observed super-threshold multiplet. 
However, it is important to note the Morse factor is discrete and is therefore not well handled by a KDE. 
Consequently, we make individual KDEs for all Morse factor values allowed for the super-threshold events based on the observation. 
A summary of the possible detected images given observed Morse factor differences is given in Table~\ref{tab:MorseFactorDiffs}. 
For a given Morse factor setup, we pick the lensed events from the catalogue respecting the SNR criterion, hence leading to the correct number of super-threshold images. 
Once this is done for all the possible Morse factor configurations, we recombine the individual KDEs by doing a weighted sum, leading to the final distribution. We can then plug in the observed time delay and relative magnification for the super-threshold events and the measured time delay for every sub-threshold trigger needing follow-up to evaluate its probability under the hypothesis that it is a lensed counterpart of the observed super-threshold images.

Let us illustrate the process with a simple example where we have observed three super-threshold images with $n_{21} = 0$ and $n_{31} = 0.5$ (4$^{\mathrm{th}}$ row in Table~\ref{tab:MorseFactorDiffs}). 
This situation has two sub-cases: one where the missing image is the fourth one in the quadruplet and one where it is the third image. 
So, we make the first KDE by selecting events in the catalogue with the three first images super-threshold, and the last one sub-threshold, according to the SNR criterion defined earlier. 
Then, we do the same for the case where the third image is sub-threshold and all the others have their SNR larger than 8. 
With the catalogue, we obtained 389 and 52 lensed events matching the first and second cases respectively.\footnote{The difference in the sample sizes is reasonable because the fourth image tends to be fainter than the third, leading to a larger sample size for the first case.}
Using these samples, we then make two KDEs representing the two probability distributions
\begin{align}
     p(t_{s1} |  \vec{\Phi}_{j1}, n_{21} = 0, n_{31} = 0.5, \mathrm{detected} &= \{1, 2, 3\}, \mathcal{H}_L, \mathcal{M}) \, , \nonumber \\
      p(t_{s1} |  \vec{\Phi}_{j1}, n_{21} = 0, n_{31} = 0.5, \mathrm{detected} &= \{1, 2, 4\}, \mathcal{H}_L, \mathcal{M}) \, . \nonumber
\end{align}
These distributions are then combined in a reweighted sum to obtain the final probability.

\begin{figure*}
  \centering

  \subfloat[%
  \label{fig:result:0}
  1 super-threshold image, no $n_{j1}$,
  no simulated events plugged in. This leads to a generic distribution that remains the same for all events.]{
    \includegraphics[width=0.48\linewidth]{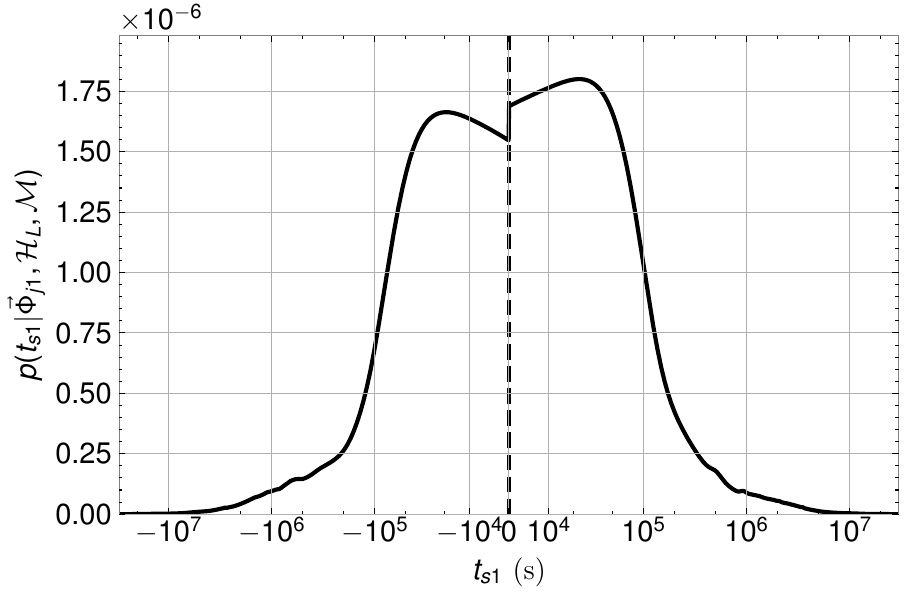}
  }
  \quad
  \subfloat[%
  \label{fig:result:123}
    2 super-threshold images, $n_{21} = 0$,
    images 1 and 2 of the simulated events are super-threshold,
    image 3 is sub-threshold]{
    \includegraphics[width=0.48\linewidth]{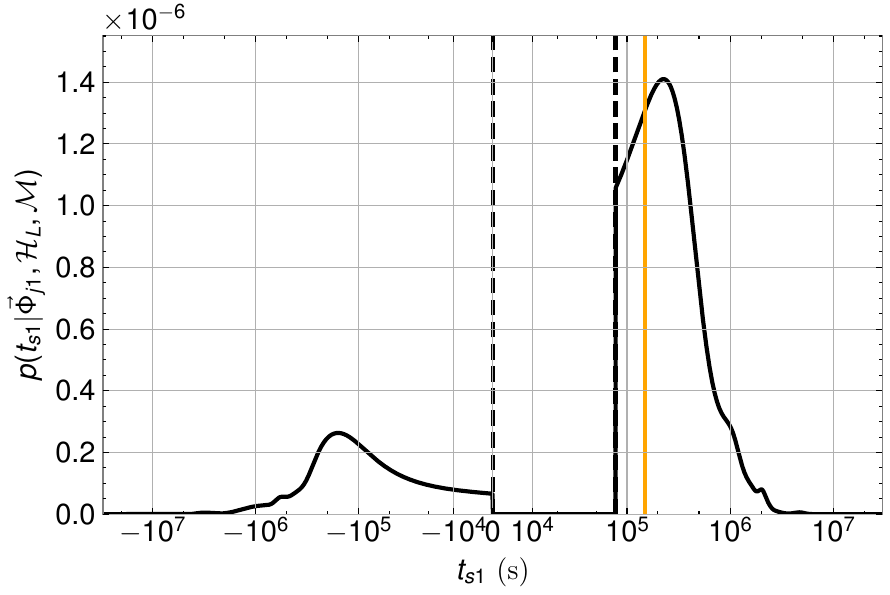}
  }
  \hfill
  \subfloat[%
  \label{fig:result:132}
  2 super-threshold images, $n_{21} = \frac{1}{2}$,
  image 1 and 3 of the simulated events are super-threshold,
  image 2 is sub-threshold]{
    \includegraphics[width=0.48\linewidth]{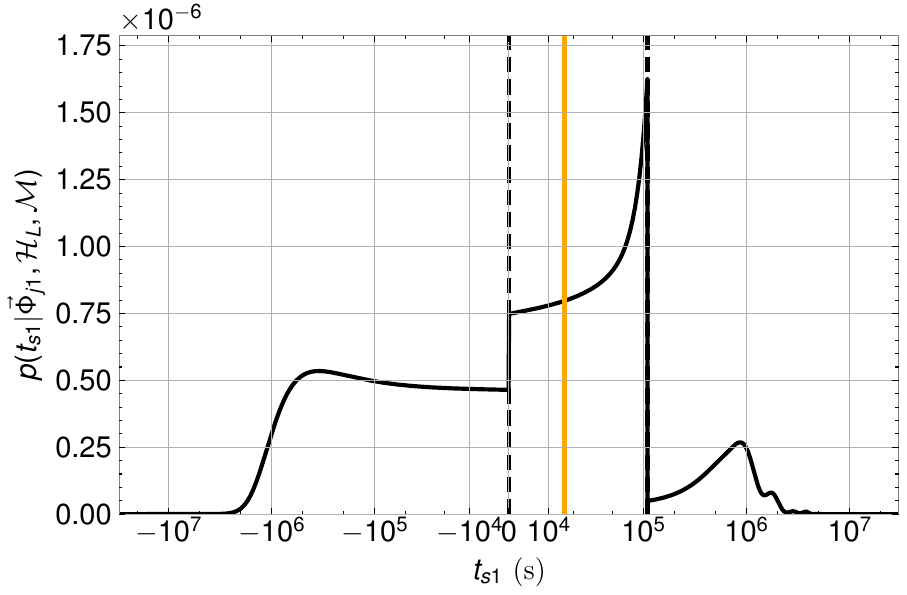}
  }
  \quad
  \subfloat[%
  \label{fig:result:231}
  2 super-threshold images, $n_{21} = \frac{1}{2}$,
  image 2 and 3 of simulated events are super-threshold,
  image 1 is sub-threshold]{
    \includegraphics[width=0.48\linewidth]{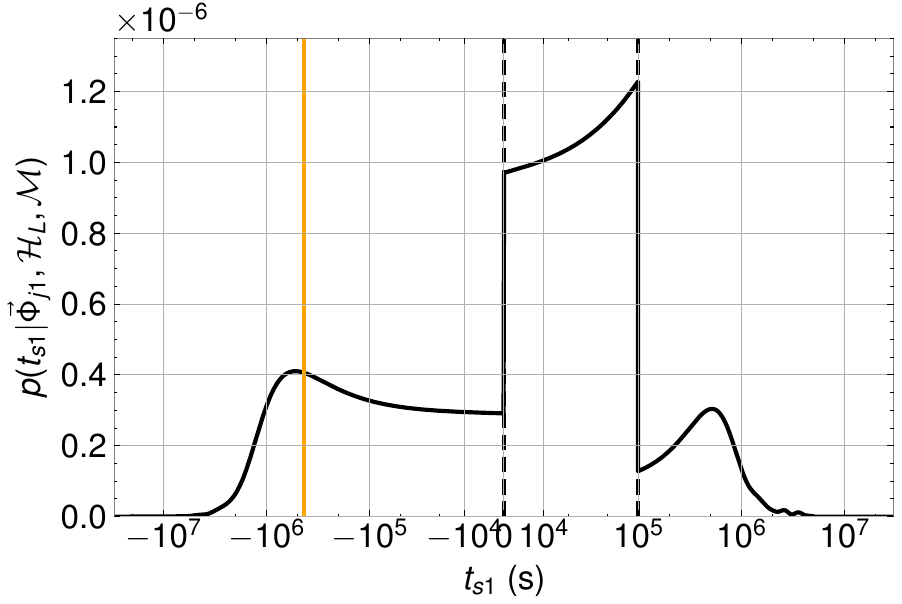}
  }
  \hfill
  \subfloat[%
  \label{fig:results:0123}
  3 super-threshold images, $n_{j1} = \{0, \frac{1}{2}\}$,
  image 1, 2, and 3 of the simulated events are super-threshold,
  image 4 is sub-threshold]{
    \includegraphics[width=0.48\linewidth]{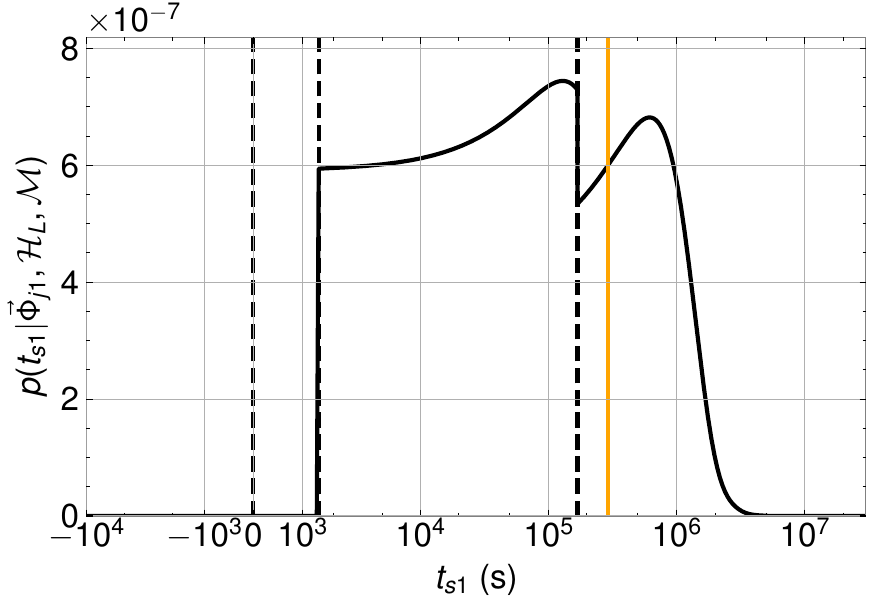}
  }
  \quad
  \subfloat[%
  \label{fig:results:1230}
  3 super-threshold images, $n_{j1} = \{\frac{1}{2}, \frac{1}{2}\}$,
  image 2, 3 and 4 of the simulated events are super-threshold,
  image 1 is sub-threshold]{
    \includegraphics[width=0.48\linewidth]{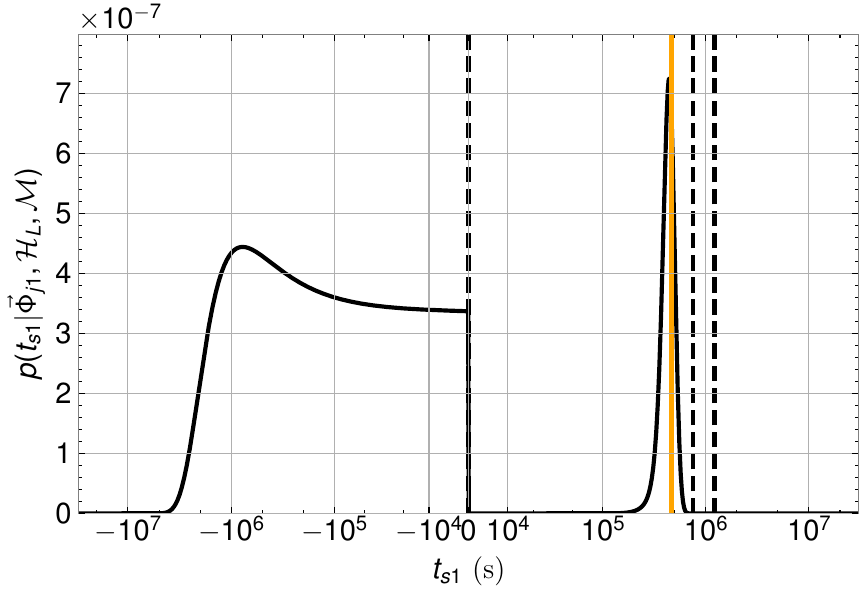}
  }
  \caption{
    $p(t_{s1} | \vec{\Phi}_{j1}, \mathcal{H}_L, \mathcal{M})$ against $t_{s1}$, for different cases of $n_{j1}$.
    For each figure, a simulated event is selected, and its true time delay is represented in orange. The dashed black lines represent the time delays between the super-threshold events relative to the first one.}
  \label{fig:dt_distributions_cases}
\end{figure*}

In the above example, it is worth noting that since the Morse factor differences are sufficient to tell we have seen the two first images, we also know that no genuine counterpart can exist before the second observed image, meaning the time delay probability is zero and we can exclude any trigger falling in that region. 
A region with a zero probability always arises when we have three images. 
When we have only two images, such a region can exist only if we have two images with the same type, as we know no image should exist in between. 
However, in this case, the region is generally shorter and the reduction in triggers to follow-up is milder. This can also be observed in Fig.~\ref{fig:dt_distributions_cases}, where we represent the final distribution obtained depending on the observed super-threshold images and covering the cases depicted in Table~\ref{tab:MorseFactorDiffs}. 
A more detailed breakdown with the various contributions of each sub-case is given in Appendix~\ref{sec:appendix_breakdown_plots}. 
In particular, Figs~\ref{fig:result:123},~\ref{fig:results:0123}, and~\ref{fig:results:1230} show the cases where we have a zero probability region. 

For reference, Fig.~\ref{fig:result:0} shows the time delay distribution when a single super-threshold image is observed. 
The probability of the sub-threshold image arriving later is larger than before. 
It is expected since the third and fourth images are more likely to be sub-threshold. 
This result is similar to that presented in other work~\citep{Wierda:2021upe, More:2021kpb, Goyal:2023lqf}. 
For all the figures, discontinuities at the super-threshold events can be observed. 
This behaviour is expected as observing a sub-threshold event before and after a super-threshold one corresponds to different situations with different probabilities. 
Figs.~\ref{fig:result:132} and ~\ref{fig:result:231}, also show that when one has two super-threshold images with different image types, \emph{i.e.} $n_{21} = 0.5$, we do not have a region with a clear zero probability, meaning no triggers will be immediately discarded. 
Still, the features of the probability distribution positively impact the re-ranking process.

\subsection{Re-Ranking the Triggers}

Above, we showed the behaviour of the time delay distribution under the lensed hypothesis. 
Here, we describe how it can be used to re-rank triggers. 
One can use an approach similar to the ones used in~\citet{Haris:2018vmn, Wierda:2021upe, More:2021kpb, Janquart_2022, Goyal:2023lqf}, where the probability under the lensed hypothesis is compared to that under the unlensed hypothesis in the ratio
\begin{equation}\label{eq:proba_ratio}
    \rgalsub = \frac{
    p(t_{s1} | \vec{\Phi}_{j1}, \lenshypo, \mods)
    }
    {
    p(t_{s1} | \vec{\Phi}_{j1}, \unlenshypo, \mods)
    } \,,
\end{equation}
where the numerator is Eq.~\eqref{eq:dt_lenshypo}, and the denominator represents the same quantity under the unlensed hypothesis, denoted by $\unlenshypo$. 
When this ratio is greater than one, it means the likelihood of observing the given time delay for the trigger is larger under the hypothesis that it is a lensed counterpart of the sub-threshold events than under the hypothesis that it is an uncorrelated signal. 
If it is smaller than one, then it is the opposite. 
We also note that when the value is close to one, $\rgalsub$ is not informative in the ranking. 

In a simpler approach, one can take the denominator in Eq.~\eqref{eq:proba_ratio} to be analytical if we assume the GW events to follow a Poisson distribution. 
However, one can get more accurate results by relying on the catalogues defined above. 
Therefore, in addition to a lens population, we construct an unlensed BBH population using the same source population and \texttt{LeR}~\citep{phurailatpam_hemantakumar_2023_8087461}. 
However, under the unlensed hypothesis, the trigger is not correlated with the lensed events, meaning it does not depend on the measured lensing parameters of the system and the probability reduces to that of the time delay between a lensed image and any sub-threshold unlensed trigger in the catalogue. 
Therefore, $p(t_{s1} | \vec{\Phi}_{j1}, \unlenshypo, \mods)$ does not require a multi-KDE procedure and we can make a single KDE representing the probability distribution for all scenarios. 
To do so, we simulate one year of unlensed events and select those with their network SNR between 6 and 8. 
Then, we also take the lensed super-threshold images and compute the time delay between them and the unlensed sub-threshold images. 
This gives us the samples from which we can make a KDE to evaluate $\rgalsub$.

Once we also have the unlensed KDE, we can compute $\rgalsub$ for all the triggers corresponding to a given set of lensed super-threshold images. 
In the cases where the lensed probability is zero, $\rgalsub = 0$, and triggers are directly discarded. 
For the others, we can rank them in descending order of $\rgalsub$, obtaining a ranked list of triggers. 
The time delay of a lower-ranked trigger is less consistent with the lensed hypothesis than the ones with a higher rank. 
Since this method is statistical, there can be cases where the lensed counterpart is in a lower probability region than some unlensed triggers. 
However, the method suggested here is complementary to other methods such as using sky map overlap between the triggers and the super-threshold images~\citep{Wong:2021lxf, Goyal:2023lqf}---which should be more efficient when considering more than one super-threshold image thanks to the reduced sky location offered by the joint observation of multiple images~\citep{Janquart:2021qov}---or using the proximity in chirp mass~\citep{Goyal:2023lqf}. 
Studying the full interplay between those methods and our approach is left for future work.

\section{Using the Time Delay Probability to Re-Rank Triggers}
\label{sec:results}

\subsection{Mock Data Setup}

In this section, we demonstrate our method's ranking capabilities on an approximate experimental setup. 
We generate 50 lensed systems with two or three super-threshold events and at least one sub-threshold event, following the population described above. 
Our testing set has 30 cases with two super-threshold images and 20 where we could observe a super-threshold triplet. 
We perform JPE on all these doubles and triples using the \textsc{GOLUM} package~\citep{Golum_git} and use their posteriors to evaluate $\rgalsub$. 
More precisely, we use the median values for the time delay and the relative magnification, and the dominant modes of the Morse factor difference.
The Morse factor differences is expected to be recovered especially confidently for genuine lensed events due to its discrete nature.

To avoid the computational burden of simulating a full year of data in which we inject each image of the lensed systems, we proceed more approximately. 
We generated 20000 unlensed events distributed over the year with \texttt{LeR} and the same BBH population models in Sec.\ref{sec:constructing_kde} and keep only the sub-threshold ones, hence with their network SNR between 6 and 8. 
This leads to 132 sub-threshold false triggers for each lensed system. 
Since we do not run the sub-threshold searches, we cannot associate the usual likelihood ratio and FAR as the ranking statistic. 
Instead, we use an approach closer to representing what would be done if some rapid follow-up has already been done by ranking the events according to their proximity in chirp mass~\citep{Goyal:2023lqf}. 
However, since we do not do template bank searches, we do not have access to the best-matching template. 
Therefore, we use the posterior overlap approach, computing the associated Bayes factor~\citep{Haris:2018vmn} using only the chirp mass posteriors, noted $\blumc$. 
Moreover, to avoid the computational burden of running parameter estimation on all the sub-threshold triggers, including the lensed counterparts and the unlensed events, we generate a mock posterior by assuming it is a Gaussian distribution with a mean equal to the true value of the chirp mass, and standard deviation equal to the mean mass divided by the SNR, following~\citet{Caliskan:2022wbh}. 
For the super-threshold images, we use the posteriors coming from JPE. 
We then use the $\blumc$ statistics for the various events as an initial ranking. 

It is important to note here that this setup is an approximation and does not translate to the result from genuine sub-threshold searches like the ones presented in~\citet{McIsaac:2019use, Li:2019osa, Li:2023tex}. 
For those, one would go through a full year of data and calculate the FAR for the various triggers, which translates to how likely it is for noise to be the root of the observed trigger. 
These searches do not fully account for the lensing hypothesis, imposing only some level of matching between the parameters through the template(s) selected to perform the sub-threshold searches. 
The method we use resembles the chirp mass proximity computation done in~\citet{Goyal:2023lqf}. 
Nevertheless, the goal here is to show the effect of ranking using $\rgalsub$, and the list obtained with this method is sufficient for this purpose. 
We leave a formal analysis of realistic data with the inclusion of $\rgalsub$ into sub-threshold search pipelines for future work and focus here on showing the main behaviours of our approach.

\subsection{Re-Ranking the Triggers}

To do the re-ranking, for every detected super-threshold lensed multiplet, we make the KDE and compute $\rgalsub$ for each sub-threshold trigger in the list. 
As the final ranking statistic, we use $\blumc \times \rgalsub$. 

\begin{figure*}
 \centering
 \subfloat[%
  \label{fig:mock:0123}
  3 super-threshold images, $n_{j1} = \{0, \frac{1}{2}\}$
  ]{
    \begin{minipage}[c]{0.48\linewidth}%
      \centering
      \includegraphics[width=1\linewidth]{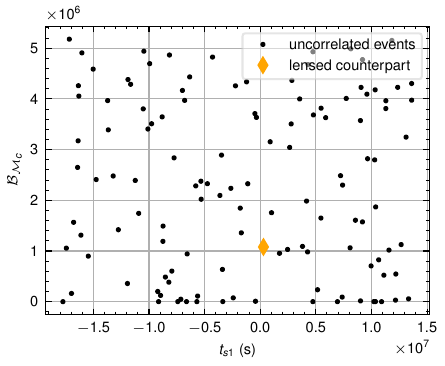} \\
      \includegraphics[width=1\linewidth]{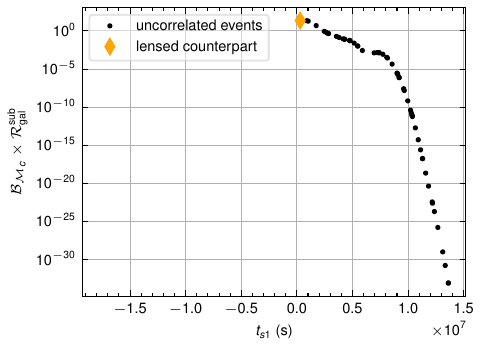}
    \end{minipage}
  }
  \subfloat[%
  \label{fig:mock:1230}
  3 super-threshold images, $n_{j1} = \{\frac{1}{2}, \frac{1}{2}\}$
  ]{
    \begin{minipage}[c]{0.48\linewidth}%
      \centering
      \includegraphics[width=1\linewidth]{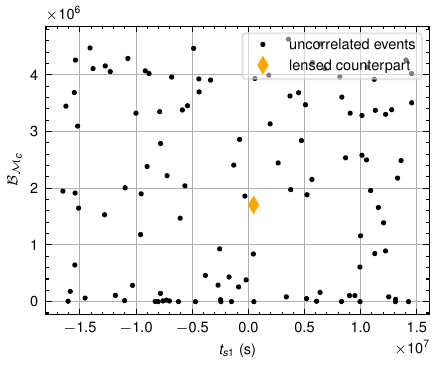} \\
      \includegraphics[width=1\linewidth]{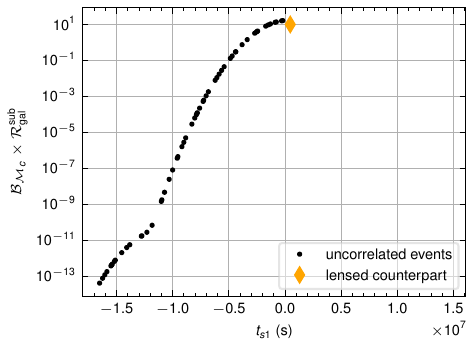}
    \end{minipage}
  }
  \caption{
    The posterior overlap statistics ($\blumc$) (top) and $\blumc\times\rgalsub$ (bottom) for the lensed sub-threshold image (orange diamond) and unrelated sub-threshold events (black dots). 
    These events are the same scenarios as the one shown in Figs.~\ref{fig:results:0123} and~\ref{fig:results:1230}, corresponding to the left and right columns, respectively. 
    We see that the presence of a null probability for some time delays discards many triggers. 
    On the left, the genuine counterpart goes from being ranked 83\textsuperscript{rd} out of 123 to being ranked 1\textsuperscript{st} out of 55, while on the right-hand side, we go from rank 65\textsuperscript{th} to 5\textsuperscript{th}, and the list goes from 113 to 62 triggers. Note that triggers may not be promoted to the first place because of the statistical nature of our approach.
    }
  \label{fig:illustrative_results}
\end{figure*}

First, in Fig.~\ref{fig:illustrative_results}, we show an illustration of our method on the same events as those shown in Figs.~\ref{fig:results:0123} and~\ref{fig:results:1230}. 
The top row corresponds to the initial ranking based only on chirp mass proximity for each event, while the second one shows the result after re-ranking. 
In both cases, the list of triggers requiring follow-up has decreased significantly since these events have a time delay region for which there is no support under the lensing hypothesis. 
Moreover, for the left-hand side, the lensed counterpart ends up with the highest rank, while it was not at the start. 
So, two main effects can be seen here: (i) the genuine lensed counterpart is better ranked after applying $\rgalsub$, and (ii) the list of triggers to follow-up is decreased. 
In the examples shown in the figure, for the left-hand side, we go from an event ranked 83\textsuperscript{rd} out of 123, to ranked 1\textsuperscript{st} out of 55. For the other case, we go from an event ranked 65\textsuperscript{th} out of 113 to one with a rank of 5\textsuperscript{th} but in a list of 62 triggers.

\begin{figure}
    \centering
    \includegraphics[width=\linewidth]{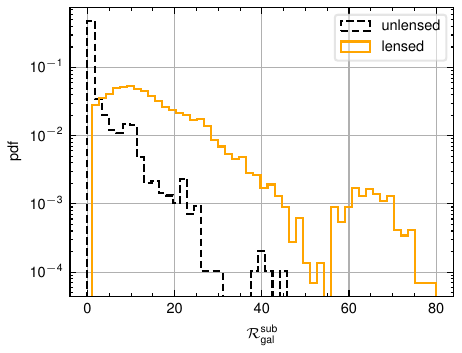}
    \caption{The expected distribution for $\rgalsub$ obtained from making a KDE of the initial distribution obtained from our samples. One sees that the lensed distribution extends much higher than the unlensed one, and, generally, lensed events have a higher ranking statistic. Note also the peak value at zero for the unlensed case, leading to many triggers being discarded. The slight bimodality in the lensed case is probably due to the relatively low number of lensed events considered in this example.}
    \label{fig:kde_distribution}
\end{figure}

\begin{figure}
    \centering
    \includegraphics[width=\linewidth]{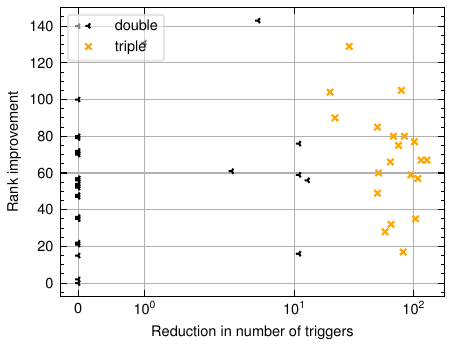}
    \caption{The improvements to the rank for our 50 lensed systems. In black are the cases for which we observe two super-threshold images, and the orange crosses are the cases where we see three events. For doubles, we generally only have a reduction in the trigger list size in some specific cases. For triples, we always have a region with zero probability for the time delay, meaning the number of triggers in the list is always significantly decreased. In all cases, we observe an improvement in rank for the genuine counterpart.}
    \label{fig:all_triples}
\end{figure}

We then carry the procedure out for our 50 lensed multiplets. 
First, in Fig.~\ref{fig:kde_distribution}, we represent the $\rgalsub$ distribution for the lensed and unlensed triggers after performing a KDE on our results. 
We see there is some overlap between the two sets of data but the statistic is generally higher for the lensed counterparts than for the unlensed events. 
Therefore, the lensed counterparts are generally expected to have an improvement in ranking when using this statistic. 
We then multiply the original ranking statistic $\blumc$ by $\rgalsub$ and use this as the final ranking for all events. 
The improvement in ranks and reduction in number of triggers in the lists requiring follow-up is shown in Fig.~\ref{fig:all_triples}. 
Some doubles do not have any reduction in the list of triggers because (a) when we see two super-threshold images with different types, there is no region with a zero probability for the time delay, and (b) even when such a region exists for doublets with the same image type, it is very short, generally spanning a few days, compared to the case when we see three super-threshold images, in which the probability of a whole region before or after a certain image is zero. 
In our examples, we have two scenarios in which the list of triggers is reduced. 
One is the case where the images have the same image type. The other is when the images have different image types but with a short time delay between them. 
This happens for low-mass lenses, which means that very long time delays are implausible and some of the triggers with the largest time delays have a negligible probability under the lensed hypothesis due to numerical errors.
Therefore, as expected, the method works better when more images are used to constrain the system. 
Nevertheless, the rank improvement is still effective in all cases considered. 

Breaking down the cases of two and three detected lensed images, we can see different effects. When two lensed super-threshold GWs are observed, the trigger list is not significantly reduced: the average reduction is 2, but the ranking of the counterpart is significantly improved. For the latter, on average, the triggers start with rank 66 and end up with rank 5, meaning they go from being in the top 48.5\% to being in the top 3.7\%. When we observe three super-threshold events, both effects become important, the rank of the genuine counterpart goes from 82 to 9 on average, while the list is reduced by a factor of 2.6, going from 133 to 52 triggers requiring follow-up. So, in this case, on average, we go from counterparts ranked in the top 61.6\% initially to them being ranked in the top 17.3\% after applying $\rgalsub$. The difference in effect comes from the difference in characteristics we can build for the different cases, with milder probability differences but regions with zero support for the triples, compared to something more variable for two images but without exclusion regions. Putting all cases together, on average, our method enables one to go from an event ranked 75 out of 133, to an event with rank 7 out of 122, meaning it is generally in the top 10 of a slightly reduced list. So, our method consistently improves the ranking of the genuine counterpart but also directly discards some of the events without requiring more expensive follow-up methods.

It is also worth noting that our approach is computationally cheap compared to running parameter estimation methods on the triggers. The heaviest part is to obtain the initial lensed catalogue from \texttt{LeR}, which takes $\mathcal{O}(hr)$\footnote{Using 8 cores on an Apple M2 Pro CPU.}. This operation needs to be done a single time per observation condition as the KDEs required to calculate $\rgalsub$ can all be derived from it. Once we have a sufficiently large catalogue, making the time delay probability distributions and computing the ranking statistics for all the triggers only takes a few minutes. 

Finally, let us also stress that, although our approach generally improves the rank of the genuine lensed counterpart and reduces the list of candidates, it does not guarantee that the event we are looking for is top-ranked. So, the safest approach if one does not want to miss the counterpart is still to go through the entire---or at least a significant fraction---of the re-ranked list with more advanced analysis methods. When one wants to do this fully, the list reduction offered by our approach also turns out to be important. Additionally, as mentioned above, our method can also be combined with other methods used to classify or veto triggers, such as imposing a sky overlap between the detected image(s) and the trigger~\citep{Wong:2021lxf} or imposing proximity in the chirp masses estimated using by the low-latency searches~\citep{Goyal:2023lqf}. Another possibility to make lensing searches more efficient is to include the lensing statistics information in the sub-threshold search likelihood itself~(Li et al., in prep). We leave the study of combining our approach with others, as well as the application of our method on realistic sub-threshold search results, for future work. 

\section{Conclusion and Future Prospects}
\label{sec:conclusions}

In this paper, we investigate the use of existing super-threshold lensed gravitational-wave detections to predict the time delays of additional sub-threshold lensed images, assuming strong galaxy lensing. 
Taking the ratio of this probability with that of the time delay under the unlensed hypothesis, one can make a ranking statistic for sub-threshold triggers translating the likelihood of them being a counterpart to the observed super-threshold lensed event. 
This re-ranking promotes the genuine counterpart in the higher density region of the lensed distribution. 
In some cases, we also expect it to reduce the list of triggers requiring follow-up since some time delays have a zero probability under the lensed hypothesis. 
This effect is due to the fixed ordering of the images as well as our capacity to measure the difference in Morse factors unambiguously in most cases.

To construct the probability distribution on a per-event basis, we use a catalogue of simulated lensed events from which we select events corresponding to our observation scenario. 
Because the Morse factors are discrete and lead to different possible individual Morse factors for the images, we make a KDE for every sub-case before combining them in a weighted sum representing the entire probability distribution. 
For the unlensed distribution, the time delay probability distribution does not depend on the characteristics of the observed lensed super-threshold images and we can make a single KDE for all cases. 
Once we have the two probability distributions for a given detected super-threshold multiplet, we can evaluate their ratio for all the sub-threshold counterpart candidates and use the result to re-rank the list.

To test the effectiveness of the method, we considered a simplified setup consisting of 50 lensed systems made of two or three super-threshold images and, at least, one sub-threshold image. 
For the super-threshold images, we ran joint parameter estimation to obtain their combined posteriors. 
To emulate an initial list of candidates, we generate sub-threshold events from a population model, construct posterior distributions following a Gaussian centred on the true values with widths scaled according to the events' SNR, and compute the posterior overlap statistics between the chirp mass posterior for the unlensed counterpart and the super-threshold images. 
While not the same as the result from a genuine sub-threshold search system, this resembles other re-ranking methods proposed in the literature to follow-up on sub-threshold candidates. 
For every system considered, we compute our ranking statistics based on the measured time delay for the sub-threshold triggers. 
The final ranking is the product between the posterior overlap and our statistics. 
When doing this, we effectively reduce the number of triggers to follow-up, with an average factor improvement of 1.2, but the trigger list is more than halved when three super-threshold images are observed. 
Moreover, we have improved rankings in all cases. 
This demonstrates our method can improve the trigger list one would pass to more expensive follow-up searches. 

We note that, in principle, if more than one super-threshold image is identified, the joint posteriors for the two first images can be used to make the template bank and evaluate the sky overlap if desired. 
While this would probably not impact much the template bank itself, the sky overlap condition would probably become more stringent since joint parameter estimation effectively reduces the observed sky area~\citep{Janquart:2021qov}. 
Moreover, sky overlap is complementary to our approach for reducing strong lensing detection candidates, and therefore using both the ranking based on sky map overlap and on the time delay compatibility would likely further reduce the list of candidates requiring follow-up. 
Finally, let us mention that if a pair of super-threshold events is detected, predicting and observing a third and/or fourth sub-threshold counterpart would strengthen the case for lensing.

More testing and development are needed to use the method in realistic scenarios. 
For example, one could improve the selection of the measured relative lensing parameters to evaluate the $\rgalsub$ ratio, accounting for their full distribution and not their median values. 
In addition, one could verify the compatibility with other ranking and vetoing methods proposed in the literature, such as the ones used in~\citet{Wong:2021lxf} and~\citet{Goyal:2023lqf}, for example. 
Ultimately, these methodologies should be tested on realistic sub-threshold search results, which is beyond the scope of this paper. 

In the end, this work shows that one can use the observed relative lensing parameters for a super-threshold lensed doublet or triplet to build a probability distribution for the time delay of a possible sub-threshold counterpart---predicting where such events should be situated in the data stream. 
Using a simplified setup, we showed that using this distribution to rank sub-threshold triggers is an effective approach because it generally promotes the genuine counterpart and reduces the list of triggers requiring more extensive follow-up. 
Finally, this method is also complementary to other ranking methods proposed in the literature, such as using the sky overlap between events. 
We hope that our work gives further motivation to the use of lensing time delay probability distributions in strong lensing searches as a mean to reduce the number of triggers requiring more expensive follow-up analyses.

\section*{Acknowledgements}
\label{sec:Acknow}
The authors thank Mick Wright, Aidan Chong, Alvin Li, and Juno Chan for useful discussions on the topic. We also thank David Keitel for useful comments on the manuscript. 
L.C.Y.N., H.P., J.S.C.P., and O.A.H. acknowledge support by grants from the Research Grants Council of Hong Kong (Project No. CUHK 14304622 and 14307923), the start-up grant from the Chinese University of Hong Kong, and the Direct Grant for Research from the Research Committee of The Chinese University of Hong Kong. 
J.J., H.N., and C.V.D.B. are supported by the research program of the Netherlands Organisation for Scientific Research (NWO).
The authors are grateful for computational resources provided by the LIGO Laboratory and supported by the National Science Foundation Grants No. PHY-0757058 and No. PHY-0823459. This material is based upon work supported by NSF’s LIGO Laboratory which is a major facility fully funded by the National Science Foundation.

\section*{Data availability}
The data underlying this article will be shared on reasonable request to the corresponding authors.

\bibliographystyle{mnras}
\bibliography{biblio}

\onecolumn

\appendix

\section{Breakdown of the Probability Distributions for Different Sub-cases}
\label{sec:appendix_breakdown_plots}
\setcounter{figure}{0} %

In this appendix, we show a case-by-case breakdown of the probability distributions for the different sub-cases explained in Sec.~\ref{sec:constructing_kde} and Table~\ref{tab:MorseFactorDiffs}. The total conditioned KDEs (same as the ones in Sec.~\ref{sec:constructing_kde}) are shown on the left, and the corresponding breakdowns are shown on the right.

\begin{center}
    \includegraphics[height=1.95in]{figures/total_conditioned_kde_singleImage.pdf}
  \quad
    \includegraphics[height=1.95in]{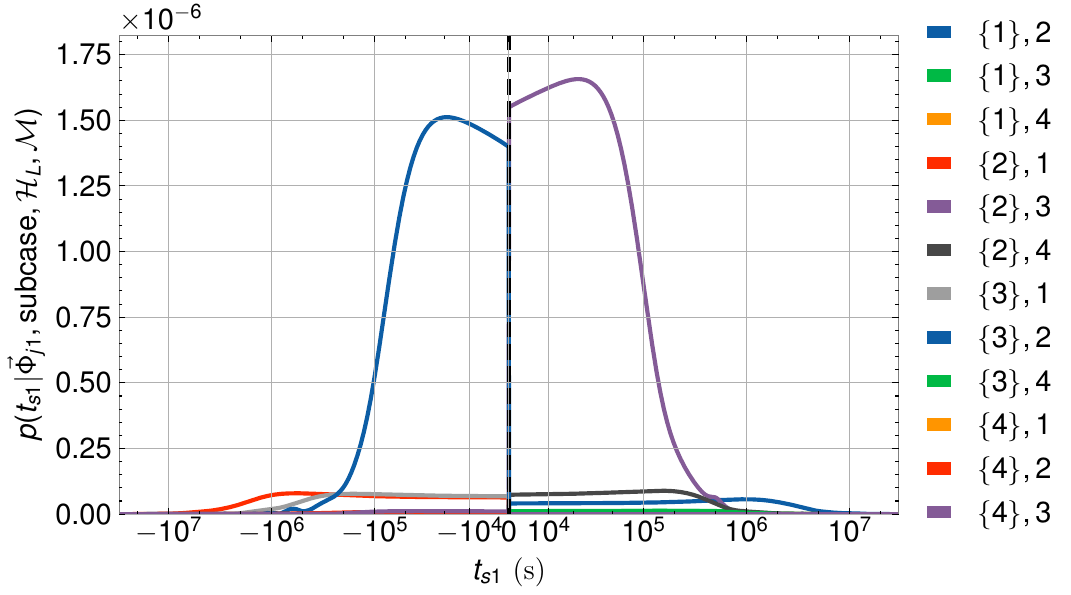}
\end{center}
\captionof{figure}{1 super-threshold image, no $n_{j1}$, last indices of legends in (b) are sub-threshold images, the rest are super-threshold images}
\vspace{1cm}

\begin{center}
    \includegraphics[height=1.95in]{figures/total_conditioned_kde_10.pdf}
  \quad
    \includegraphics[height=1.95in]{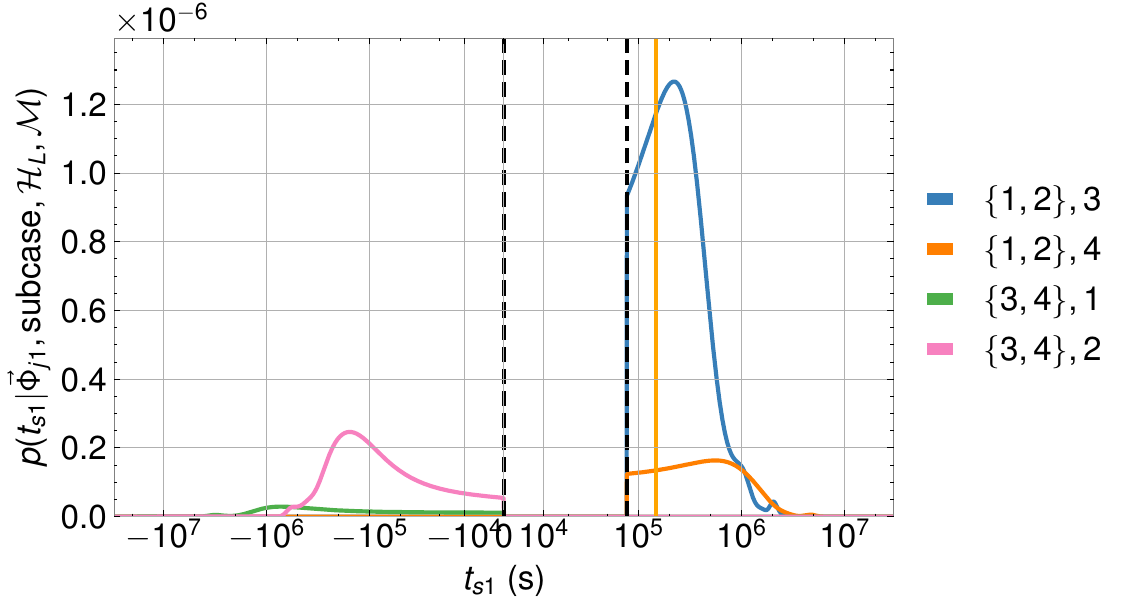}
\end{center}
\captionof{figure}{2 super-threshold images, $n_{21} = 0$, last indices of legends in (b) are sub-threshold images, the rest are super-threshold images}
\vspace{1cm}

\begin{center}
    \includegraphics[height=1.95in]{figures/total_conditioned_kde_4.pdf}
  \quad
    \includegraphics[height=1.95in]{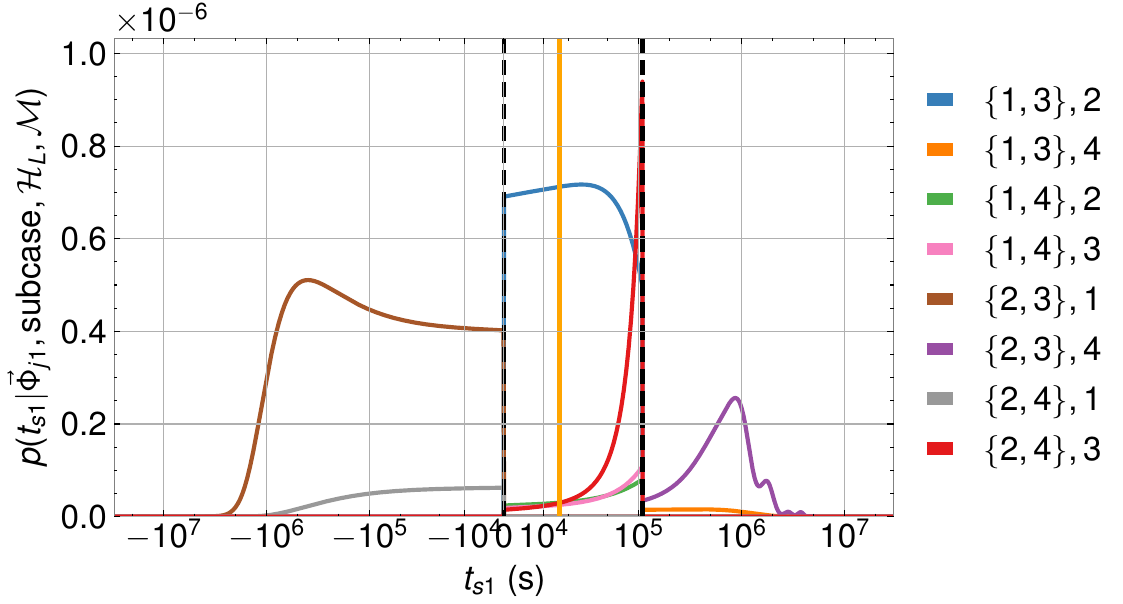}
\end{center}
\captionof{figure}{2 super-threshold images, $n_{21} = \frac{1}{2}$, last indices of legends in (b) are sub-threshold images, the rest are super-threshold images}
\vspace{1cm}

\newpage

\begin{center}
    \includegraphics[height=1.95in]{figures/total_conditioned_kde_23.pdf}
  \quad
    \includegraphics[height=1.95in]{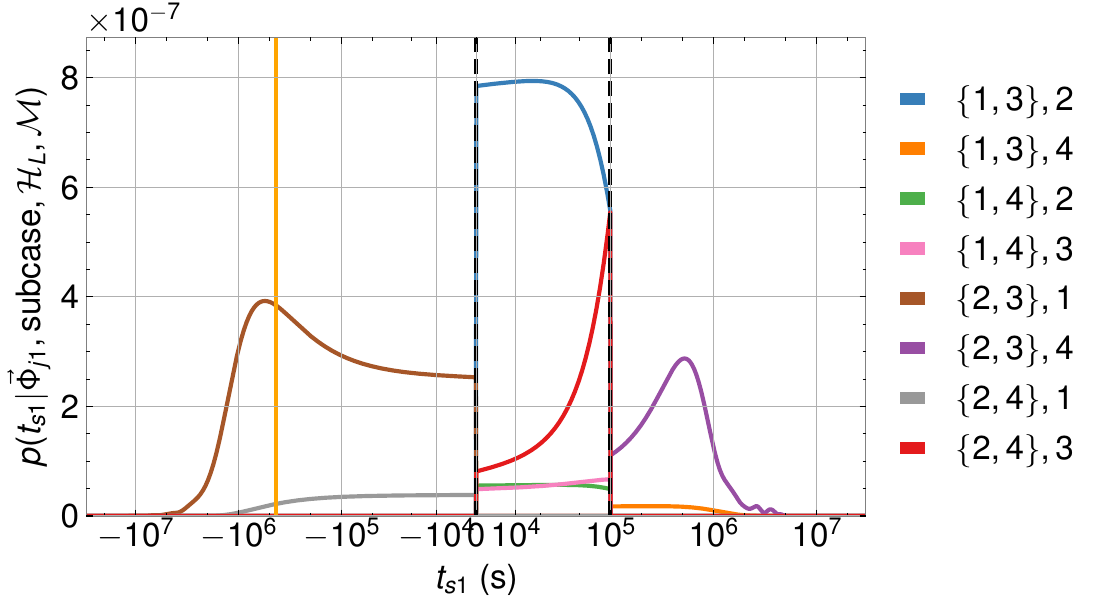}
\end{center}
\captionof{figure}{2 super-threshold images, $n_{21} = \frac{1}{2}$, last indices of legends in (b) are sub-threshold images, the rest are super-threshold images}
\vspace{1cm}

\begin{center}
    \includegraphics[height=1.95in]{figures/total_conditioned_kde_1.pdf}
  \quad
    \includegraphics[height=1.95in]{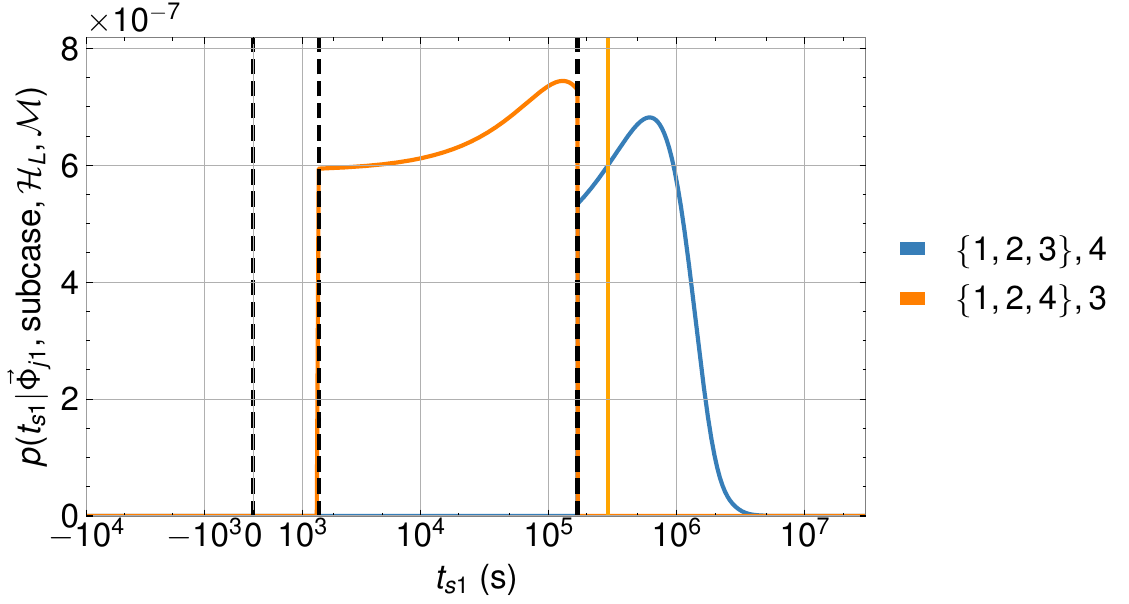}
\end{center}
\captionof{figure}{3 super-threshold images, $n_{j1} = \{0, \frac{1}{2}\}$, last indices of legends in (b) are sub-threshold images, the rest are super-threshold images}
\vspace{1cm}

\begin{center}
    \includegraphics[height=1.95in]{figures/total_conditioned_kde_32.pdf}
  \quad
    \includegraphics[height=1.95in]{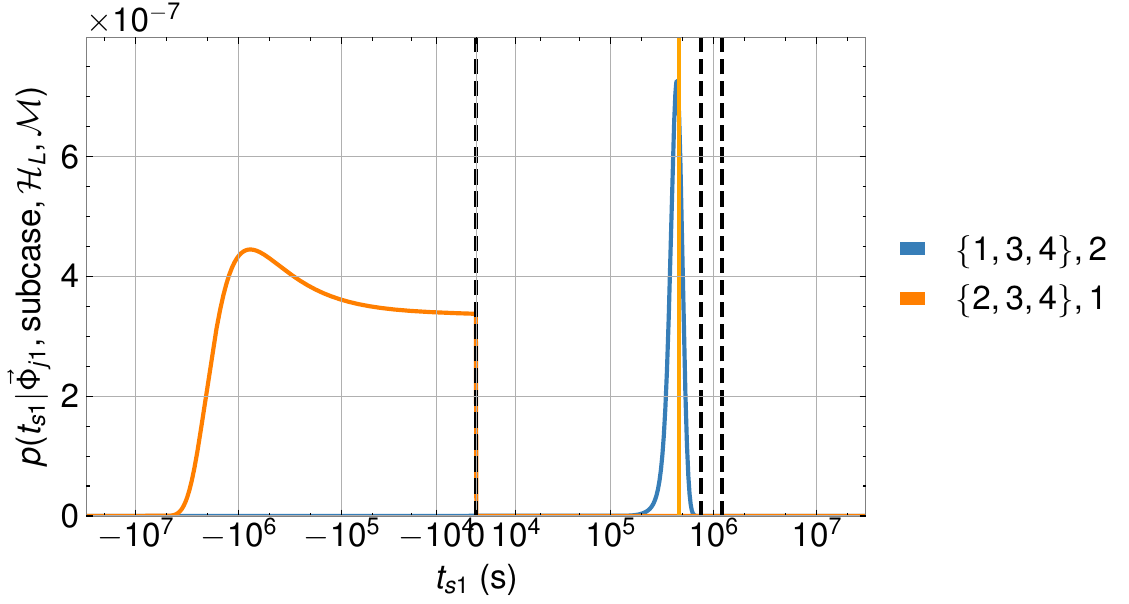}
\end{center}
\captionof{figure}{3 super-threshold images, $n_{j1} = \{\frac{1}{2}, \frac{1}{2}\}$, last indices of legends in (b) are sub-threshold images, the rest are super-threshold images}
\vspace{1cm}
%%%

\newpage
\section{Parametes of models used for simulation}
\label{appenB}
\setcounter{table}{0}

The simulation was done by first sampling the lenses and BBHs from existing models. The sampling methods and the models are detailed in the documentation of \href{https://ler.readthedocs.io/en/latest/Lensed_events.html#Sampling-Lens-Properties}{ler}. The parameters used to sample the BBHs population are listed below.

\begin{center}
  \centering
  \begin{tabular}{|l|l|}
    \hline
    Parameter & Value \\
    \hline
        $\lambda_{\text{peak}}$ & $0.10$ \\
        $\alpha$ & $2.63$ \\
        $\beta$ & $1.26$ \\
        $\mu_m$ & $33.07 \ M_\odot$ \\
        $\sigma_m$ & $5.69 \ M_\odot$ \\
        $m_{\text{max}}$ & $86.22 \ M_\odot$ \\
        $m_{\text{min}}$ & $4.59 \ M_\odot$ \\
        $\delta_m$ & $4.82 \ M_\odot$ \\
    \hline
  \end{tabular}
  \label{tab:BBH_mass_distributions_parameters}
\captionof{table}{BBH mass distributions model parameters, based on the GWTC-2 Population results~\citep{2021arXiv211103634T}.}
\end{center}

\begin{center}
  \centering
  \begin{tabular}{|l|l|}
    \hline
    Parameter & Value \\
    \hline
        $\lambda_{z}$ & $0.563$ \\
        $a$ & $2.906$ \\
        $b$ & $0.0158$ \\
        $c$ & $0.58$ \\
        $\mu_z$ & $1.1375$ \\
        $\sigma_z$ & $0.8665$ \\
    \hline
  \end{tabular}
  \label{tab:BBH_redshift_distributions_parameters}
\captionof{table}{Parameters for $d V_c / d z$, the differential comoving volume~\citep{Wierda:2021upe}. This gives the redshift distribution of BBHs.}
\end{center}

\end{document}